\documentclass{article}
\usepackage[utf8]{inputenc}
\usepackage{tikz}
\usepackage{multirow}
\usepackage{amsfonts}
\usepackage{amsmath}
\usepackage{amssymb}
\usepackage{subfigure}
\usepackage{url}
\usepackage{algorithm2e}
\usepackage{garamond}
\usepackage[urw-garamond]{mathdesign}
\newenvironment{keywords}{
       \list{}{\advance\topsep by0.35cm\relax\small
       \leftmargin=1cm
       \labelwidth=0.35cm
       \listparindent=0.35cm
       \itemindent\listparindent
       \rightmargin\leftmargin}\item[\hskip\labelsep
                                     \bfseries Keywords:]}
     {\endlist}
\def\similar{\textsc{SIMILAR}}
\def\irms{\textsc{IRM4S}}
\def\irmmls{\textsc{IRM4MLS}}
\def\padawan{\textsc{PADAWAN}}
\def\gama{\textsc{GAMA}}

\newcommand\strong[{1}]{\textbf{#1}}

\title{On time and consistency in multi-level agent-based simulations}

\author{%
	Gildas MORVAN, Yoann KUBERA\\
{\small \url{http://www.lgi2a.univ-artois.fr/~morvan/}}\\{\small gildas.morvan@univ-artois.fr}\\~\\Univ. Artois, EA 3926,\\Laboratoire de Génie Informatique et
    d’Automatique de l’Artois (LGI2A)\\Béthune, France\\~\\
\includegraphics[width=3cm]{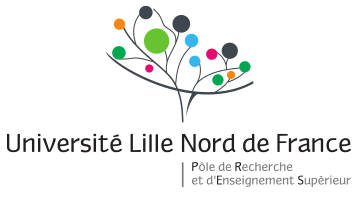}~~~\includegraphics[width=3cm]{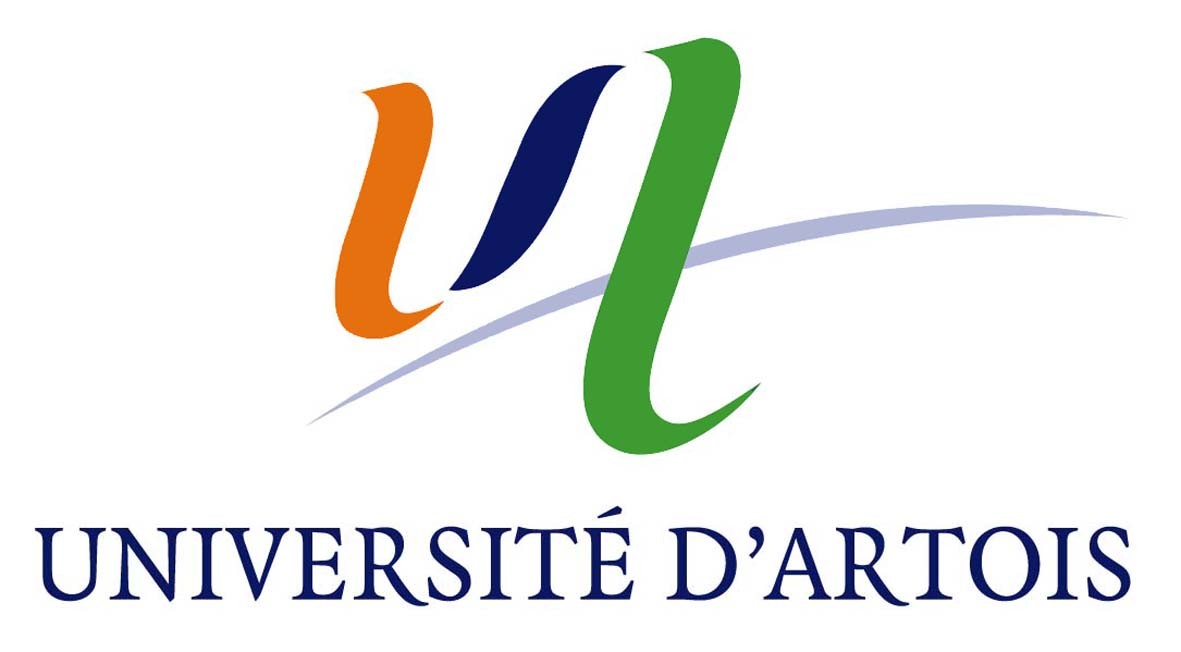}~~~\includegraphics[width=2.5cm]{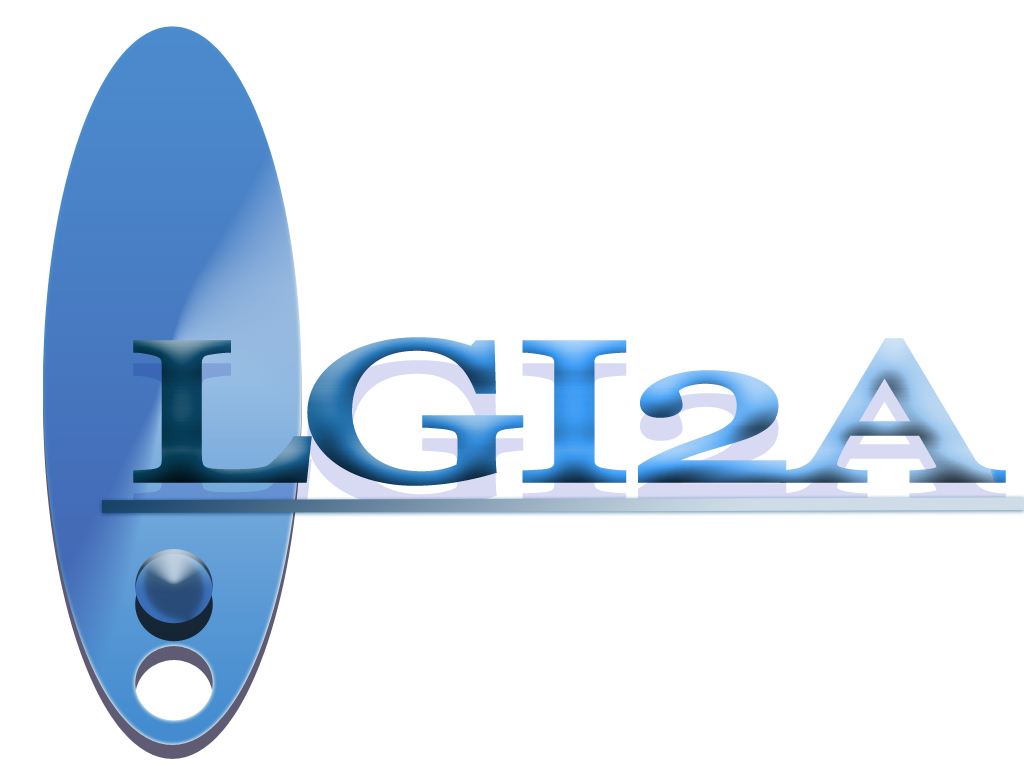}
}

\date{}
\begin{document}

\maketitle

\begin{abstract}
The integration of multiple viewpoints became an increasingly popular approach to deal with 
agent-based simulations. Despite their disparities, recent approaches successfully manage to 
run such multi-level simulations. Yet, are they doing it appropriately?

This paper tries to answer that question, with an analysis based on a generic model of the 
temporal dynamics of multi-level simulations.
This generic model is then used to build an orthogonal approach to multi-level simulation called 
\similar{}. In this approach, most time-related issues are explicitly modeled, owing to an 
implementation-oriented approach based on the influence/reaction principle.
\end{abstract}


\section*{Introduction}
Simulating complex systems often requires the integration of knowledge coming from 
different viewpoints (\textit{e.g.} different application fields, different focus points) 
to obtain relevant results.
Yet, the representations of the agents, the environment and the temporal dynamics in regular multi-agent 
based simulation meta-models are designed to support a single viewpoint.
Therefore, they lack the structure to manage the integration of such systems, called 
\emph{multi-level simulations}.

Managing multiple viewpoints on the same phenomenon induces the use of heterogeneous time models, thus 
raising issues related to \emph{time} and \emph{consistency}.
Multi-Level Agent-based Modeling (ML-ABM) is a recent approach that aims at extending the classical 
single-viewpoint agent-based modeling paradigm to cope with these issues and create multiple-viewpoints 
based simulations~\cite{Camus:2015a,David:2009,Gil-Quijano:2012,Hjorth:2015b,Huraux:2014,Morvan:2013,Picault:2011,Vo:2012b}.
Considering the disparities between the various ML-ABM approaches, a natural question comes to mind: 
\emph{is there a "right" way to do ML-ABM?}

In this context, the aim of this paper is double. We first aim at eliciting the issues and simulation
choices underlying such simulations, with an analysis based on a generic model of the temporal dynamics 
of a multi-level simulation. Then, we present \similar{}, a ML-ABM approach using the 
influence/reaction principle to manage explicitly the issues related to the simultaneous actions of 
 agents in multiple levels~\cite{Ferber:1996,Michel:2007,Morvan:2011}.

\section{\label{sec:temporal dyn}Temporal dynamics in Multi-level simulations}
In this section some issues related to multi-level 
simulation are emphasized using a generic model describing the temporal dynamics
of a multi-level simulation.
%%
%% %% %% %% %% %% %% %% %% %% %% %% %% %% %% %% %% %% %% %% %% %% %% %% %% %% %% %%
%% SUBSECTION
%% %% %% %% %% %% %% %% %% %% %% %% %% %% %% %% %% %% %% %% %% %% %% %% %% %% %% %%
%%
\subsection{General case}
From a coarse grain viewpoint, simulation is a process transforming the 
data about a phenomenon from initial values into a sequence of intermediate 
values, until a final state is reached. This 
evolution is characterized by:
\strong{1)}  A \emph{dynamic state} $\delta{}(t) \in \Delta$ modeling the data of the 
	simulation at time $t$;
	\strong{2)}  A \emph{time model} $\mathbb{T}$ representing the moments when each state 
	of the discrete evolution was obtained;
	\strong{3)}  A \emph{behavior model} describing the evolution process of the dynamic 
	state between two consecutive moments of the time model.
	
The exact content of the time model, dynamic state, as well as the behavior model of a simulation 
 depends on the simulation approach being used. Yet, despite their disparities, many common points 
can be identified among them.

First, since real time can be seen as a continuous value, most simulation assume that 
$\mathbb{T}\subset\mathbb{R}$. Moreover, we can assume that a simulation eventually ends. 
Thus, $\mathbb{T}$ contains an ordered, finite and discrete set of time values $t\in\mathbb{T}$. 
%The construction of this set will depend on the multi-level approach being used.
Then, the dynamic state contains data 
related to the \emph{agents} and the \emph{environment}\footnote{In this paper, we use a simplistic definition of these concepts: an agent is an 
entity that can \emph{perceive} data about itself, the environment and the other agents, 
possibly \emph{memorize} some of them and \emph{decide} to perform actions}.
%%
%% %% %% %% %% %% %% %% %% %% %% %% %% %% %% %% %% %% %% %% %% %% %% %% %% %% %% %%
%% SUBSECTION
%% %% %% %% %% %% %% %% %% %% %% %% %% %% %% %% %% %% %% %% %% %% %% %% %% %% %% %%
%%
\subsection{Multi-level case}
In multi-level simulations, each level embodies a specific viewpoint on the studied phenomenon. Since 
these viewpoints can evolve using very different time scales, each level $l\in\mathbb{L}$ (where $
\mathbb{L}$ is the set of all levels) has to define its own time model $\mathbb{T}_l$.

The  interaction of the levels is possible only by defining when and under which circumstances 
interaction is possible. For this purpose, we introduce a multi-level specific 
terminology to the temporal dynamics.
%%
%% %% %% %% %% %% %% %% %% %% %% %% %% %% %% %% %% %% %% %% %% %% %% %% %% %% %% %%
%% SUBSUBSECTION
%% %% %% %% %% %% %% %% %% %% %% %% %% %% %% %% %% %% %% %% %% %% %% %% %% %% %% %%
%%
\subsubsection{Local information}
We consider that agents can lie in more than one level at a time.
$\mathcal{A}(t,l)\in\mathbb{A}_l$ denotes the set of agents of the level 
$l\in\mathbb{L}$ at time $t\in\mathbb{T}_l$. 
Since levels can have very different temporal dynamics, this point has various
implications on the structure of the simulation:
\strong{1)} Agents have a \emph{local state}\footnote{Also called "physical state" or "face" in the literature~\cite{Ferber:1996,Michel:2007,Morvan:2011,Picault:2011}} $\phi_{a}(t, l)\in\Phi_{a,l}$ in each level $l\in\mathcal{L}$ where they lie;
\strong{2)} Agents perform decisions differently depending on the level from which the decisions  originates;
\strong{3)} A level $l$ can only trigger the local behavior of the agents lying in $l$.

Similarly, the environment has a local state $\phi_{\omega}(t,l)\in\Phi_\omega$ in each level of 
the simulation. Yet, contrary to the agents, the environment is present in each level of the 
simulation. Each local state embodies any agent-independent information like a topology or a state 
(\textit{e.g.} an ambient temperature).
%%
%% %% %% %% %% %% %% %% %% %% %% %% %% %% %% %% %% %% %% %% %% %% %% %% %% %% %% %%
%% SUBSUBSECTION
%% %% %% %% %% %% %% %% %% %% %% %% %% %% %% %% %% %% %% %% %% %% %% %% %% %% %% %%
%%
\subsubsection{Global information}
The coherence of agent behaviors in each level can require information like 
cross-level plans or any other level-independent information. Therefore, 
we consider that agents have a \emph{global state}\footnote{Also called "mind", 
"memory state" or "core" in the literature~\cite{Ferber:1996,Michel:2007,Morvan:2011,Picault:2011}} $\mu_a(t)\in\mathbb{M}_a$, which is independent 
from any level.
%%
%% %% %% %% %% %% %% %% %% %% %% %% %% %% %% %% %% %% %% %% %% %% %% %% %% %% %% %%
%% SUBSUBSECTION
%% %% %% %% %% %% %% %% %% %% %% %% %% %% %% %% %% %% %% %% %% %% %% %% %% %% %% %%
%%
\subsubsection{Content of a dynamic state}
Owing to the abovementioned information, the dynamic state $\delta(t)\in\Delta$ of a multi-level
simulation at time $t$ can be defined as the sum of the \emph{local dynamic state} 
$\delta(t,l)\in\Delta_l$ of each level $l\in\mathbb{L}$ and the \emph{global dynamic state} 
$\delta_G(t)\in\Delta_G$ containing the global state of the agents.

\begin{equation}
\Delta = \Delta_G \times \prod_{l\in\mathbb{L}}\Delta_l
\end{equation}

\begin{equation}
\forall \delta (t) \in \Delta, \delta (t) = \Big( \delta_G (t), \big( \delta(t,l) \big)_{{l\in
\mathbb{L}}} \Big) \mbox{ with } \delta_G(t)\in\Delta_G \wedge \forall l\in\mathbb{L}, \delta(t,l) 
\in \Delta_l
\end{equation}

%%
%% %% %% %% %% %% %% %% %% %% %% %% %% %% %% %% %% %% %% %% %% %% %% %% %% %% %% %%
%% SUBSUBSECTION
%% %% %% %% %% %% %% %% %% %% %% %% %% %% %% %% %% %% %% %% %% %% %% %% %% %% %% %%
%%
\subsubsection{Time model of the simulation}
The interaction between levels is possible only if their time models are somehow 
correlated.
Since the time model of each level is a discrete ordered set, it is possible to build 
an order between their elements.

The time model of a multi-level simulation is defined as the union of the time models
of all the levels: $\mathbb{T} = \bigcup_{l\in\mathbb{L}} \mathbb{T}_l$. For consistency reasons,
the time models $\mathbb{T}$ and $\mathbb{T}_l$ must have the same bounds. Since $\mathbb{T}$ and 
$\mathbb{T}_l$ are ordered, we also define $s+dt$ (resp. $s+dt_l$) as the successor of 
$s\in\mathbb{T}$ (resp. $s\in\mathbb{T}_l$).
%%
%% %% %% %% %% %% %% %% %% %% %% %% %% %% %% %% %% %% %% %% %% %% %% %% %% %% %% %%
%% SUBSUBSECTION
%% %% %% %% %% %% %% %% %% %% %% %% %% %% %% %% %% %% %% %% %% %% %% %% %% %% %% %%
%%
\subsubsection{Consistent and transitory states}
In the case where $t\not\in\mathbb{T}_l$, the level $l\in\mathbb{L}$ is in a 
\emph{transitory state}. No guarantee can be provided on such a state, since it corresponds to a temporary value 
used by $l$ to compute its future consistent dynamic state.
On the opposite, the data contained in the dynamic state of a 
level $l\in\mathbb{L}$ can be safely read or perceived at times in $\mathbb{T}_l$, 
where this state is considered as \emph{consistent}. 
\begin{figure}[htbp]
	\includegraphics[width=\textwidth]{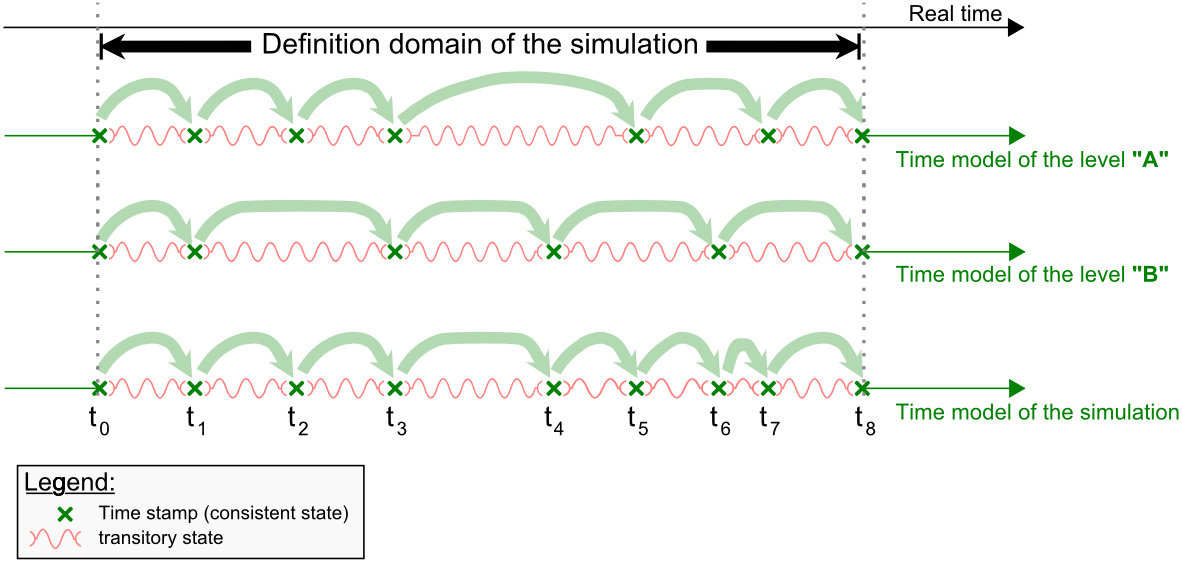}
	\caption{\label{fig:time model}%
		Illustration of a time model, for a simulation using two levels "A" %
		and "B".%
		The first line represents real time.%
		The second and third line represent the time model of the levels "A" 
		and "B". %
		The last line represents the time model of the whole simulation. %
		At $t_1$, the state of the simulation is consistent. At $t_4$, it is %
		half-consistent.
	}
\end{figure}

From a global viewpoint, the dynamic state of the simulation is 
\emph{consistent} (resp. \emph{transitory}) if all of its levels are consistent (resp. 
transitory). It can also be in an intermediate situation called \emph{half-consistent 
state}, if a level is in a consistent and another level is in a transitory state. 
These concepts are illustrated in Figure~\ref{fig:time model}.

To clarify our speech, we write $\delta{}(t)$ the \emph{consistent (or 
half-consistent) dynamic state} of the simulation at time $t\in\mathbb{T}$ and 
$\delta(]t,t^\prime[)$ the \emph{transitory dynamic state} of the simulation 
between the times $t\in\mathbb{T}$ and $t^\prime\in\mathbb{T}$. 
%%
%% %% %% %% %% %% %% %% %% %% %% %% %% %% %% %% %% %% %% %% %% %% %% %% %% %% %% %%
%% SUBSECTION
%% %% %% %% %% %% %% %% %% %% %% %% %% %% %% %% %% %% %% %% %% %% %% %% %% %% %% %%
%%
\subsection{Multi-level inherent issues}
When a simulation is in a transitory phase, each level performs operations \emph{in parallel}
to determine the next consistent value of their dynamic state. The transitory periods of the 
levels are not necessarily in sync. Therefore, each level can be at a different step of its
transitory operations when an interaction occurs. This point raises the 
following time-related issues: \strong{1)} Determine on which dynamic states is based the decision in a level to interact 
	with another level; \strong{2)} Determine when to take into consideration the modifications in a level resulting 
	from an action initiated in another level; \strong{3)} Determine how to preserve the consistency of the global state of the agents 
	despite having its update occurring after the level-dependent perceptions.

\begin{figure}[htbp]
	\includegraphics[width=\textwidth]{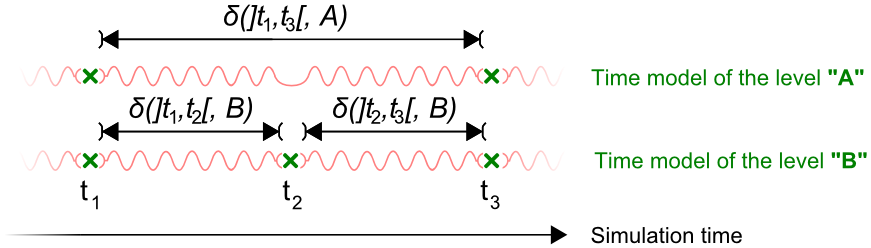}
	\caption{\label{fig:ml issue}%
		Illustration of a subset of a time model containing two levels "A" and "B".
	}
\end{figure}

This section illustrates these issues on an example containing two levels "A" and "B", presented in 
Figure~\ref{fig:ml issue} 
%%
%% %% %% %% %% %% %% %% %% %% %% %% %% %% %% %% %% %% %% %% %% %% %% %% %% %% %% %%
%% SUBSUBSECTION
%% %% %% %% %% %% %% %% %% %% %% %% %% %% %% %% %% %% %% %% %% %% %% %% %% %% %% %%
%%
\subsubsection{Level interaction through perception}
The first issue is related to the perception of the dynamic state of the other levels. It happens 
for instance at the time $t_2$ (see fig.~\ref{fig:ml issue}), when an agent from the level "B" tries to read information from the level "A". 
Indeed, since "A" is in a transitory state at that time, the data being read by the
agent might have arbitrary values. Therefore, a heuristic has to be used to \emph{disambiguate} 
that value. For instance using the last consistent dynamic state of the level (in this case the 
dynamic state of "A" at the time $t_1$), using the arbitrary values from the transitory state at 
$t_2$ or anticipating the modifications that might have occurred in "A" between $t_1$ and $t_2$.
%%
%% %% %% %% %% %% %% %% %% %% %% %% %% %% %% %% %% %% %% %% %% %% %% %% %% %% %% %%
%% SUBSUBSECTION
%% %% %% %% %% %% %% %% %% %% %% %% %% %% %% %% %% %% %% %% %% %% %% %% %% %% %% %%
%%
\subsubsection{Level interaction through actions}
The second issue is related to the side effects of an interaction between two levels. It happens for
instance during the transitory period $]t_1,t_3[$  (see fig.~\ref{fig:ml issue}) of the level 
"A", if an agent from "A" tries to interact with the level "B". Indeed, since both levels have
a different time scale, it is difficult to determine when the actions of "A" have to be taken for
account into the computation of the dynamic state of "B". It can be during the transitory phases
$]t_1,t_2[$, $]t_2,t_3[$ or a later one.

A generic answer to that problem might be "the next time both levels are in sync" ($t_3$ in this 
case). Yet, this leads to aberrations like taking into consideration these actions at the end 
of the simulation (for instance in fig~\ref{fig:time model}, if an agent from the level "A" interacts 
with "B" during the transitory period $]t_3,t_5[$).
%%
%% %% %% %% %% %% %% %% %% %% %% %% %% %% %% %% %% %% %% %% %% %% %% %% %% %% %% %%
%% SUBSUBSECTION
%% %% %% %% %% %% %% %% %% %% %% %% %% %% %% %% %% %% %% %% %% %% %% %% %% %% %% %%
%%
\subsubsection{Global state update}
The third issue is related to the read and write access of the global state of an agent and the 
update of that value.
Indeed, during the transitory phase of each level, the agent has to read and possibly update the 
value of the global states, to take into account the information that were perceived. Yet, since the 
perception is relative to each level, the global state is the subject of the same issues than the 
interactions between levels.

For instance, in figure~\ref{fig:ml issue}, the period $]t_1,t_2[$ of the simulation is a 
transitory period for the level "B" and a subset of the transitory period for the level "A".
The latter raises the question of whether if the global state of the agent at the time $t_2$ has 
to take into consideration the data being perceived by the agent from "A" or not. Indeed, perception
might not be complete at that time in "A".
%%
%% %% %% %% %% %% %% %% %% %% %% %% %% %% %% %% %% %% %% %% %% %% %% %% %% %% %% %%
%% SUBSECTION
%% %% %% %% %% %% %% %% %% %% %% %% %% %% %% %% %% %% %% %% %% %% %% %% %% %% %% %%
%%
\subsection{\label{sec:diff btw ml appr}Differences between multi-level approaches}
There is no universal answer to the issues presented in this section, since the coherence 
between heterogeneous time scales is itself an ill-defined notion. The main differences between
existing ML-ABM approaches are the way these issues are handled, through the answer of
the following questions about the operations performed during a transitory phase $]t,t+dt[$:
\strong{1)} Which agents can perform a decision during a transitory period of a level?
\strong{2)} How many actions can be performed by the decision process of an agent?
\strong{3)} How are committed the results of the action to the future dynamic state of a level?
\strong{4)} When are performed these operations during the transitory state?
\strong{5)} Which dynamic state of a level $k$ is read by a level $l$ initiating an interaction with $k$? A consistent one? A transitory one? Which ones ?
\strong{6)} When is taken into account the interaction initiated by a level $l$ with a level $k$?
\strong{7)} How is managed the consistency of the global state of agents?

In the next section, we present an agent-based approach called 
\similar{}, that aims at addressing these issues.

\section{\label{sec:similar}SIMILAR}
Many meta-models and simulation engines dedicated to ML-ABM have been proposed in the 
literature such as \irmmls{}~\cite{Morvan:2011}, \padawan{}~\cite{Picault:2011},  
\gama{}~\cite{Grignard:2013} or \textsc{NetLogo LevelSpace}~\cite{Hjorth:2015b}. 
All these approaches provide a different and yet valid answer to the multi-level 
simulation issues.
In this paper, we do not aim at detailing precisely their differences: a comprehensive 
survey of the different approaches can be found in~\cite{Morvan:2013}.

Existing approaches like \gama{} or \padawan{} (Pattern for Accurate Design of Agent 
Worlds in Agent Nests) are complete approaches providing various interesting features 
respectively including the agentification of emerging structures or the elicitation of 
interactions between agents. 
However, these approaches rely on a time model where the management of the potentially 
simultaneous actions is strongly constrained by the sequential execution of agent 
actions.

In this paper, we investigate another approach where agent actions are separated from 
their consequences in the dynamic state of the simulation, using the 
influence/reaction principle~\cite{Ferber:1996}.
The resulting approach, called \similar{} (\textsc{Si}mulations with 
\textsc{M}ult\textsc{i}-\textsc{L}evel \textsc{A}gents and \textsc{R}eactions), is 
deeply inspired by \irmmls{}~\cite{Morvan:2011}, a multi-level extension of 
\irms{}~\cite{Michel:2007}.
The main differences between \similar{} and \irmmls{} are the more precise and less 
misleading terminology and simulation algorithms, as well as a more precise and 
implementation-oriented model for the reaction phase (the latter is not described in 
this paper due to the lack of space).
%%
%% %% %% %% %% %% %% %% %% %% %% %% %% %% %% %% %% %% %% %% %% %% %% %% %% %% %% %%
%% SUBSECTION
%% %% %% %% %% %% %% %% %% %% %% %% %% %% %% %% %% %% %% %% %% %% %% %% %% %% %% %%
%%
\subsection{Core concepts}
\similar{} revolves around five core concepts:
\strong{1)}  \emph{Levels}, modeling different viewpoints on the simulated phenomenon;
	\strong{2)}  \emph{Agents} lying in one or more levels. From each level where they lie, they perceive 
	the state of one or more levels to decide how they wish to influence 
	the evolution of the system;
	\strong{3)}  the \emph{Environment} modeling the topology, the local information (\emph{e.g.} 
	temperature) and the natural evolution\footnote{\textit{i.e.} without the intervention of the 
	behavior of an agent} of each level;
	\strong{4)}  \emph{Influences} modeling actions which effect has yet to be committed to the state of 
	the simulation;
	\strong{5)}  \emph{Reactions} modeling how the changes depicted by the influences are committed to 
	the state of the simulation.

We note $\mathbb{L}$ the levels defined for a simulation, $\mathbb{I}$ the domain 
space of all the possible influences of the simulation and $\mathbb{A}$ all possible agents 
of the simulation.
%%
%% %% %% %% %% %% %% %% %% %% %% %% %% %% %% %% %% %% %% %% %% %% %% %% %% %% %% %%
%% SUBSECTION
%% %% %% %% %% %% %% %% %% %% %% %% %% %% %% %% %% %% %% %% %% %% %% %% %% %% %% %%
%%
\subsection{Heuristics}
\begin{figure}[htbp]
	\includegraphics[width=\textwidth]{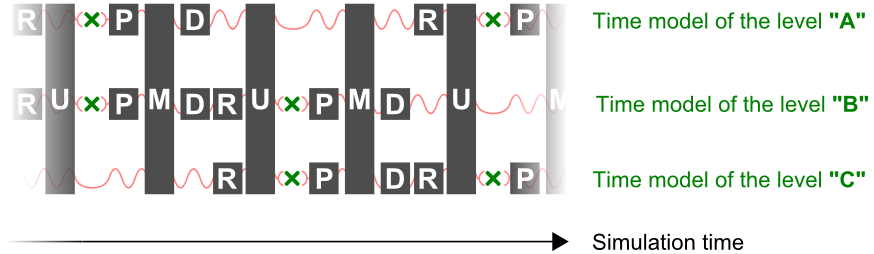}
	\caption{\label{fig:memory revision time}%
		Illustration of the operation performed independently in each level (squares) and 
		joint for all the levels (rectangles) during a simulation. The letters describe the type 
		of the operation: Perception (P), Global state revision (M), Decision (D) and Reaction (R). 
		Any arrow starting on a consistent dynamic state at a time $t$ points to the revised global 
		state used as the new global state of the agents in the half-consistent dynamic state of the 
		time $t$.
		This example focuses on the operations performed in a 
		simulation containing one agent lying in three levels.
	}
\end{figure}

\similar{} relies on the following heuristics and choices to manage the issues raised in the 
section~\ref{sec:diff btw ml appr}. %The consequences on the execution flow of the simulation is illustrated on fig.~\ref{fig:memory revision time}:
\strong{1)} During the transitory period $]t,t+dt_l[$ of a level $l\in\mathbb{L}$, the agents from  $\mathcal{A}(t,l)$ decide once in parallel;
\strong{2)} The number of influences produced by each decision is not constrained;
\strong{3)}  The result of the actions is committed to the future dynamic state of a level using a  reaction mechanism~\cite{Ferber:1996} ;
\strong{4)} During the transitory period $]t,t+dt_l[$ of a level $l\in\mathbb{L}$, the behavior of the agents is triggered slightly after $t$ and the reaction occurs slightly before $t+dt_l$;
\strong{5)} The dynamic state being read by the behavior of an agent (or of the environment) is  always the most recent consistent state of the level\footnote{Default heuristic of \similar{}. \similar{} also allows the definition of user-defined disambiguation heuristics};
\strong{6)} The actions emitted by an agent from a level $l$ to a level $k$ during a transitory  period $]t,t+dt_l[$ are taken into account in the next reaction of $k$ after the time $t$ (\textit{i.e.} the reaction occurring during the transitory period containing or starting with the time $t$);
\strong{7)} The consistency of agent global states  is attained by: \strong{i)} Computing the revised global state of the agents at the beginning of the transitory 
		period of a level (\textit{i.e.} before any reaction);
		\strong{ii)} Computing the revised global state of an agent once for all the levels 
		starting a new transitory period at the same time;
		\strong{iii)} Use this revised global state as the global state of the agent for the next half-
		consistent state of the simulation. This approach is summarized in Figure~\ref{fig:memory revision time}

%%
%% %% %% %% %% %% %% %% %% %% %% %% %% %% %% %% %% %% %% %% %% %% %% %% %% %% %% %%
%% SUBSECTION
%% %% %% %% %% %% %% %% %% %% %% %% %% %% %% %% %% %% %% %% %% %% %% %% %% %% %% %%
%%
\subsection{Dynamic state}
In \similar{}, we consider that each point of view on a phenomenon has to be 
embodied in a \emph{level} $l\in\mathbb{L}$. As a consequence, the dynamic state $\delta(t)$ of 
the simulation is divided in level-specific dynamic states $\delta(t,l)$. Two kind of data can be 
obtained from the dynamic state of a level $l\in\mathbb{L}$:  a \emph{state valuation} $\sigma{}(t, l)$, defining a valuation of the 
level-related properties of the agents (\textit{e.g.} their location or their temperature) or 
the environment (\textit{e.g.} an ambient temperature)
and the \emph{state dynamics} $\gamma{}(t, l)$, defining the actions that were 
still being performed\footnote{Actions that started before the time $t$ and that 
will end after the time $t$} in that level during the observation.

\begin{equation}
	\forall l\in\mathbb{L}, \forall t\in\mathbb{T}_l, 
	\delta{}(t, l) = < \sigma{}(t, l), \gamma{}(t, l) >
\end{equation}
%%
%% %% %% %% %% %% %% %% %% %% %% %% %% %% %% %% %% %% %% %% %% %% %% %% %% %% %% %%
%% SUBSUBSECTION
%% %% %% %% %% %% %% %% %% %% %% %% %% %% %% %% %% %% %% %% %% %% %% %% %% %% %% %%
%%
\subsubsection{State valuation}
The \emph{state valuation} $\sigma{}(t, l)$ of a level $l\in\mathbb{L}$ is the union of  the local state of the environment $\phi_{\omega} (t,l)\in\Phi_\omega$, containing 
	agent-unrelated information and  a local state\footnote{This replaces the term \emph{"physical state"} from \irmmls{}, 
	which was misleading, since that state also contains mental information like a desired 
	speed.} $\phi_a(t,l)\in\Phi_{a,l}$ for each agent $a\in\mathbb{A}$ 
	contained in the level. %These states contain information like a position, a speed or a carried item;

%%
%% %% %% %% %% %% %% %% %% %% %% %% %% %% %% %% %% %% %% %% %% %% %% %% %% %% %% %%
%% SUBSUBSECTION
%% %% %% %% %% %% %% %% %% %% %% %% %% %% %% %% %% %% %% %% %% %% %% %% %% %% %% %%
%%
\subsubsection{State dynamics}
\similar{} relies on the influence/reaction principle to model 
the actions resulting from the decision of the agents, from the natural evolution of
the environment and the actions still being performed at time $t$. Therefore, 
the \emph{state dynamics} $\gamma(t, l)$ of a level $l\in\mathbb{L}$ contains a set 
of influences.
\begin{equation}
	\forall l \in \mathbb{L}, \forall t\in\mathbb{T}_l, \gamma(t,l) \subseteq \mathbb{I}
\end{equation}
Since the data contained in an influence are mostly domain-dependent, no specific model
is attached to them. They usually contain the subjects of the action (\textit{e.g.} the 
physical state of one or more agents) as well as parameters (\textit{e.g.} an amount 
of money to exchange).
%% 
%% %% %% %% %% %% %% %% %% %% %% %% %% %% %% %% %% %% %% %% %% %% %% %% %% %% %% %%
%% SUBSECTION
%% %% %% %% %% %% %% %% %% %% %% %% %% %% %% %% %% %% %% %% %% %% %% %% %% %% %% %%
%%
\subsection{General behavior model}
The dynamic state of a simulation models a "photograph" of the simulation at time $t$. 
Motion is attained owing to the \emph{behavior of the agents}, the \emph{natural 
action of the environment} and the \emph{reaction of each level} to influences.

%\begin{figure}[htbp]
%	\includegraphics[width=\textwidth]{images/executionSingleLevel.png}
%	\caption{\label{fig:execution single level}%
%		The main operations performed during the transitory phase $t+dt_l$ of a 
%		level $l$, to move its dynamic state from the value $\delta(t,l)$ to its
%		value $\delta(t+dt_l,l)$. This figure illustrates the case where the level 
%		contains two agents $a_1$ and $a_2$.
%	}
%\end{figure}
%%
%% %% %% %% %% %% %% %% %% %% %% %% %% %% %% %% %% %% %% %% %% %% %% %% %% %% %% %%
%% SUBSUBSECTION
%% %% %% %% %% %% %% %% %% %% %% %% %% %% %% %% %% %% %% %% %% %% %% %% %% %% %% %%
%%
\subsubsection{Behavior of the agents}
The behavior of an agent in a level $l\in\mathbb{L}$ has three phases: 
\strong{1)} \emph{Perception:} extract information from the dynamic state of  the levels that can be perceived from $l$; 
\strong{2)} \emph{Global state revision:} use the newly perceived data to revise the content of the global state of the agent; 
\strong{3)} \emph{Decision:} use the perceived data and the revised global state to create and send influences to the levels that can be influenced by $l$.  Each influence models a modification request of the dynamic state of a level.
	
%%
%% %% %% %% %% %% %% %% %% %% %% %% %% %% %% %% %% %% %% %% %% %% %% %% %% %% %% %%
%% SUBSUBSECTION
%% %% %% %% %% %% %% %% %% %% %% %% %% %% %% %% %% %% %% %% %% %% %% %% %% %% %% %%
%%
\subsubsection{Natural action of the environment}
The natural action of the environment is simpler than the behavior of agents: 
it only has one phase, where the dynamic state of the levels that can be perceived 
from $l$ are used to create influences sent to one or more levels that can be influenced 
by $l$.

%For instance perceive 1)the body temperature of all the agents and the ambient 
%temperature of the environment in a "room" level; 2)the ambient temperature in 
%the "outdoors" level, in order to 3)send an "modification of the ambient temperature 
%of the environment" influence to the "room" level.
%%
%% %% %% %% %% %% %% %% %% %% %% %% %% %% %% %% %% %% %% %% %% %% %% %% %% %% %% %%
%% SUBSUBSECTION
%% %% %% %% %% %% %% %% %% %% %% %% %% %% %% %% %% %% %% %% %% %% %% %% %% %% %% %%
%%
\subsubsection{Reaction to the influences}
As in \irmmls{}, in \similar{} the reaction of a level $l\in\mathbb{L}$ is a process
computing the new consistent dynamic state of $l$. The reaction phase occurs at the
end of a transitory period $]t,t+dt_l[$ of a level, and is computed using the value of the most recent consistent dynamic state of $l$ and the influences that were sent to $l$ during the transitory period $]t,t+dt_l[$;

Yet, contrary to \irmmls{}, \similar{} provides an explicit model to the generic influences
that can be found in any simulation, like the addition/removal of an agent from the 
simulation/a level. Such influences are called \emph{system influences}, in opposition to 
\emph{regular influences}, which are user-defined. A model is also provided to their generic 
reaction. These points are not detailed in this paper. %Note that this paper provides an overview on \similar{}: the actual formalism and its implementation as a simulation library rely on a more detailed model. 
%% 
%% %% %% %% %% %% %% %% %% %% %% %% %% %% %% %% %% %% %% %% %% %% %% %% %% %% %% %%
%% SUBSECTION
%% %% %% %% %% %% %% %% %% %% %% %% %% %% %% %% %% %% %% %% %% %% %% %% %% %% %% %%
%%
\subsection{Formal notations and simulation algorithm}
Not all levels are able to interact. Therefore, the interactions between levels are constrained 
by two digraphs: 
A \emph{perception} relation graph $\mathbb{G}_P$ (resp. \emph{influence} relation 
graph $\mathbb{G}_I$) defines which levels can be perceived (resp. influenced) during the behavior 
of the agent/environment in a specific level.
\begin{equation}
	\begin{array}{lcl}
		\big( l_1,l_2 \big) \in \mathbb{G}_P \mbox{ (resp. }	\mathbb{G}_I\mbox{)} 
			& \iff 
			& \mbox{An agent from } l_1 \mbox{ can perceive (resp.}
		\\
			&
			& \mbox{ influence) the dynamic state of } l_2
	\end{array}
\end{equation}

The out neighborhood $\mathcal{N}_P^+(l)$ (resp. $\mathcal{N}_I^+(l)$) of a level 
$l\in\mathbb{L}$ in the perception (resp. influence) relation graph defines the levels
 that can be perceived (resp. influenced) by $l$.
%%
%% %% %% %% %% %% %% %% %% %% %% %% %% %% %% %% %% %% %% %% %% %% %% %% %% %% %% %%
%% SUBSUBSECTION
%% %% %% %% %% %% %% %% %% %% %% %% %% %% %% %% %% %% %% %% %% %% %% %% %% %% %% %%
%%
\subsubsection{Agent behavior}
Since the content of the dynamic state is not trustworthy during transitory periods, the 
natural action of the environment and the perception of the agents are based on the 
\emph{last consistent dynamic state} of the perceptible levels. This time is identified 
by the notation $floor_l( ]t,t^\prime[ )$, which models the last time when the dynamic 
state of a level $l$ was consistent for a perception occurring during a transitory period 
$]t,t^\prime[$.
\begin{equation}
	\forall t,t^\prime \in \mathbb{T}^2, \forall l\in\mathbb{L}, 
	floor_l( ]t,t^\prime[ ) = max\Big(\{ u \leq t | u\in\mathbb{T}_l \}\Big)
\end{equation}
Based on these information, the perception phase of an agent $a\in\mathcal{A}(t,l)$ from a level 
$l\in\mathbb{L}$ for the transitory period $]t,t+dt_l[$ is defined as an 
application $perception_{a,]t,t+dt_l[,l}$. This application reads the last consistent dynamic 
state of each perceptible level to produce the perceived data:
\begin{equation}
	\begin{array}[b]{p{1cm}lcl}
		
		\multicolumn{4}{l}{
			\forall l\in\mathbb{L}, \forall t\in\mathbb{T}_l\backslash\{max(\mathbb{T}_l)\}, 
			\forall a\in\mathcal{A}(t,l),\;\;\;\;perception_{a,]t,t+dt_l[,l}:}\\
		&
			\prod_{ k\in\mathcal{N}_P^+(l) }\Delta_k & 
			\longrightarrow & 
			\mathbb{P}_{a,l}\\
		&
			\Big( \delta\big( floor_k( ]t,t+dt_l[ ), k \big) \Big)_{k\in\mathcal{N}_P^+(l)} & 
			\longmapsto & 
			p_{a,l}(]t,t+dt_l[)
	\end{array}
\end{equation}
In this notation, $\mathbb{P}_a$ models the domain space of the data that can be perceived by
the agent $a$, from the perspective of the level $l$. It can contain raw data from the dynamic
states, or an interpretation of these data. For instance, in a road traffic simulation, the drivers 
do not need to put the absolute position of the leading vehicle (\textit{i.e.} raw data from the 
dynamic state) in their perceived data: the distance between the two vehicles is sufficient.

The revision of the global state of an agent $a\in\mathcal{A}(t,l)$ for a transitory period 
starting at the time $t$ is defined as an application $globalRev_{a,]t,t+dt[}$. This application 
reads the most recent consistent global state $\mu_a\big( t )$ of the agent $a$ and the perceived 
data $p_{a,l}(]t,t+dt_l[), \forall l\in\mathbb{L}|t\in\mathbb{T}_l$ of all the levels that 
started a transitory phase at the time $t$, in order to determine the value of the revised global 
state $\mu_a(]t,t+dt[)$ of the agent during the transitory period.
\begin{equation}
	\begin{array}[b]{p{1cm}lcl}
		\multicolumn{4}{l}{
			\forall t\in\mathbb{T}\backslash\{max(\mathbb{T})\}, 
			\forall a\in\mathcal{A}(t),\;\;\;\;globalRev_{a,]t,t+dt[}:}\\
		&
			\mathbb{M}_a \times \prod_{l\in\mathbb{L}|t\in\mathbb{T}_l}\mathbb{P}_{a,l} & 
			\longrightarrow & 
			\mathbb{M}_a\\
		&
			\Big( \mu_a( t ), \big(p_{a,l}(]t,t+dt_l[)\big)_{l\in\mathbb{L}|t\in\mathbb{T}_l} \Big) & 
			\longmapsto & 
			\mu_a( ]t,t+dt[ )
	\end{array}
\end{equation}

Finally, the decision of an agent $a\in\mathcal{A}(t,l)$ from a level 
$l\in\mathbb{L}$ for the transitory period $]t,t+dt_l[$ is defined as an 
application $decision_{a,]t,t+dt_l[,l}$. This application reads the revised global state
$\mu_a(]t,t+dt[)$ of $a$ and the perceived data $p_{a,l}(]t,t+dt_l[)$ computed for the level $l$
to create the influences that will modify levels during their respective next reaction.
\begin{equation}
	\begin{array}[b]{p{1cm}lcl}
		\multicolumn{4}{l}{
			\forall l\in\mathbb{L}, \forall t\in\mathbb{T}_l\backslash\{max(\mathbb{T}_l)\}, 
			\forall a\in\mathcal{A}(t,l),\;\;\;\;decision_{a,]t,t+dt_l[,l}:}\\
		&
			\mathbb{M}_a \times \mathbb{P}_{a,l} & 
			\longrightarrow & 
			2^\mathbb{I}\\
		&
			\Big( \mu_a( ]t,t+dt[ ), p_{a,l}( ]t,t+dt_l[ ) \Big) & 
			\longmapsto & 
			\mathcal{I}_{a,l}( ]t,t+dt_l[ )
	\end{array}
\end{equation}
If we note $level( i )$ the level at which the influence $i\in\mathbb{I}$ is aimed, then the 
influence relation graph imposes the following constraint to the decision:
\begin{equation}
	\forall l\in\mathbb{L}, \forall t\in\mathbb{T}_l\backslash\{max(\mathbb{T}_l)\}, 
	i\in\mathcal{I}_{a,l}( ]t,t+dt_l[ ) \Rightarrow level(i) \in \mathcal{N}_I^+(l)
\end{equation}
As a result to this phase, each created influence $i\in\mathcal{I}_{a,l}( ]t,t+dt_l[ )$ is added 
to the transitory state dynamics of $k=level(i)$, for the transitory period 
$]floor_{k}( ]t,t+dt_l[), floor_{k}( ]t,t+dt_l[) + dt_{k}[$.
%%
%% %% %% %% %% %% %% %% %% %% %% %% %% %% %% %% %% %% %% %% %% %% %% %% %% %% %% %%
%% SUBSUBSECTION
%% %% %% %% %% %% %% %% %% %% %% %% %% %% %% %% %% %% %% %% %% %% %% %% %% %% %% %%
%%
\subsubsection{Natural action of the environment}
The natural action of the environment from a level $l\in\mathbb{L}$ for the transitory 
period $]t,t+dt_l[$ is defined as an application $natural_{]t,t+dt_l[,l}$. 
This application reads the last consistent dynamic state of each perceptible level  
to create the influences that will modify the dynamic state of the influenceable levels 
(during their reaction).
\begin{equation}
	\begin{array}[b]{p{1cm}lcl}
		\multicolumn{4}{l}{
			\forall l\in\mathbb{L}, \forall t\in\mathbb{T}_l\backslash\{max(\mathbb{T}_l)\}, 
			\;\;\;\;natural_{]t,t+dt_l[,l}:}\\
		&
			\prod_{ k\in\mathcal{N}_P^+(l) }\Delta_k & 
			\longrightarrow & 
			2^\mathbb{I}\\
		&
			\Big( \delta\big( floor_k( ]t,t+dt_l[ ), k \big) \Big)_{k\in\mathcal{N}_P^+(l)} & 
			\longmapsto & 
			\mathcal{I}_{\omega,l}( ]t,t+dt_l[ )
	\end{array}
\end{equation}
The resulting influences are managed with the same process than the ones coming from the decisions of the agents.

%%
%% %% %% %% %% %% %% %% %% %% %% %% %% %% %% %% %% %% %% %% %% %% %% %% %% %% %% %%
%% SUBSUBSECTION
%% %% %% %% %% %% %% %% %% %% %% %% %% %% %% %% %% %% %% %% %% %% %% %% %% %% %% %%
%%
\subsubsection{Reaction of a level}
The reaction of a level $l\in\mathbb{L}$ is computed at the end of each transitory period 
$]t,t+dt_l[$ where $t\in\mathbb{T}_l$. It is defined as an application 
$reaction_{l, ]t, t+dt_l[}$ reading the transitory dynamic state
$\delta(]t,t+dt_l[, l)$ of the level to determine the next consistent value of the dynamic state 
$\delta(t+dt_l, l)$.
\begin{equation}
	\begin{array}[b]{p{1cm}lcl}
		\multicolumn{4}{l}{
			\forall l\in\mathbb{L}, \forall t\in\mathbb{T}_l\backslash\{max(\mathbb{T}_l)\}, 
			\;\;\;\;reaction_{]t,t+dt_l[,l}:}\\
		&
			\Delta_l & 
			\longrightarrow & 
			\Delta_l\\
		&
			\Big( \delta\big( ]t,t+dt_l[, l \big) \Big) & 
			\longmapsto & 
			\delta\big( t+dt_l, l \big)
	\end{array}
\end{equation}
The reaction has the following responsibilities: 
\strong{1)} Take into consideration the influences of $\gamma(]t,t+dt_l[, l)$ to update the local state of the agents, update the local state of the environment, create/delete  agents from the simulation or add/remove agents from the level;
\strong{2)} Determine if the influences of $\gamma(]t,t+dt_l[, l)$ persist in $\gamma(t+dt_l, l)$  (if they model something that has not finished at the time $t+dt_l$);
\strong{3)} Manage the colliding influences of $\gamma(]t,t+dt_l[, l)$.

%In \similar{}, the reaction of a level also includes a generic part managing the reaction to the 
%system influences. The integration of this point to the reaction is not detailed in this paper.
%%
%% %% %% %% %% %% %% %% %% %% %% %% %% %% %% %% %% %% %% %% %% %% %% %% %% %% %% %%
%% SUBSECTION
%% %% %% %% %% %% %% %% %% %% %% %% %% %% %% %% %% %% %% %% %% %% %% %% %% %% %% %%
%%
\subsubsection{Simulation algorithm}
The simulation algorithm of \similar{} is presented in 
Figure~\ref{fig:simulation algorithm}. It relies on the presented concepts and complies with the 
time constraints defined in~\cite{Morvan:2011}.

\begin{figure}[htb]
	\input{images/algo}
	\caption{\label{fig:simulation algorithm} The simulation algorithm used in \similar{}}
\end{figure}

\section*{Conclusion and perspectives}

In this paper, we elicited several issues about time and consistency raised in multi-level simulations. 
There is no clear solution to theme since the notion of time consistency 
among heterogeneous time models is itself ill-defined. Therefore, rather than distinguishing 
the "right" or "wrong" approaches, we defined a theoretical frame giving a better understanding
of the choices underlying each approach. Then, it is up to modelers and domain specialists to tell if these choices are appropriate
or not for the study of a given phenomenon.

To cope with the multi-level related issues, we introduced a meta-model named \similar{} based on the 
influence/reaction principle.
This model is designed to reify as much as possible the concepts involved in the abovementionned 
issues, thus providing a better support to the definition of explicit solutions to them.
\similar{} includes a generic and modular formal model, a methodology and a simulation API preserving 
the structure of the formal model. 
Thus, the design of simulations is in addition more robust to model revisions and relies on a structure 
fit to represent the intrinsic complexity of the simulated multi-level phenomena.

\similar{} has been implemented in Java and is available under the CeCILL-B license.

It is available at \url{http://www.lgi2a.univ-artois.fr/~morvan/similar.html}. 

\bibliographystyle{plain}
\bibliography{../../Biblio}

\end{document}